\documentclass[%
 aip,
 amsmath,amssymb,
 reprint,%
]{revtex4-1}

\usepackage{graphicx}
\usepackage{dcolumn}
\usepackage{bm}
\usepackage[mathlines]{lineno}

\usepackage[utf8]{inputenc}
\usepackage[T1]{fontenc}
\usepackage{mathptmx}
\usepackage{etoolbox}
\usepackage{amsmath}
\usepackage{upgreek}
\usepackage{xcolor}
\makeatother

\begin{document}

\preprint{AIP/123-QED}

\title{Evolutionary dynamics of collective-risk social dilemma game with risk uncertainty and adaptive strategy}

\author{Chuanzhi Zhou$^\dagger$}
\author{Ju Han$^\dagger$}

\author{Xiaojie Chen}\thanks{Corresponding author: xiaojiechen@uestc.edu.cn. \\ $^\dagger$Chuanzhi Zhou and Ju Han contributed equally to this work.}

\affiliation{School of Mathematical Sciences, University of Electronic Science and Technology of China, Chengdu 611731, China}

\author{Attila Szolnoki}

\affiliation{Institute of Technical Physics and Materials Science, Centre for Energy Research, P.O. Box 49, Budapest, H-1525, Hungary}

\begin{abstract}
The collective-risk dilemma game constitutes the leading metaphor for studying the evolution of cooperation in populations of selfish individuals. Although most theoretical studies often consider the risk level as a fixed parameter, this may not fully reflect the uncertain nature of risk detected in real-world scenarios. Motivated by the observation that individuals can infer their game environment and accordingly adjust their actual behavior, here we investigate this game with uncertain risk. In particular, we introduce a type of adaptive agents who incur an observation cost to infer the risk level during game interactions, and subsequently decide whether to cooperate or defect based on their tolerance thresholds. We first calculate the fixation probabilities of three different strategies in a mutation-free regime. Our results show that adaptive agents are favored by natural selection at low observation costs and moderate tolerance thresholds. Moreover, the fixation of adaptive agents results in higher levels of cooperation when both parameters are relatively low. We then consider behavioral mutation and systematically explore how average cooperation level and group achievement vary with observation costs, tolerance thresholds, and mutation rates. Importantly, the main results remain intact when the mutation is considered. Furthermore, when the tolerance threshold is small, the presence of adaptive agents results in a higher cooperation level and group achievement compared to the traditional setup when only unconditional cooperators and defectors are present. This advantage, however, gradually erodes with an increasing tolerance threshold and these two quantities fall below the levels observed in the classic two-strategy case.
\end{abstract}

\maketitle

\begin{quotation}
Preventing catastrophic events such as irreversible climate change requires collective action, yet participants often hesitate to cooperate due to conflict of interest. The collective-risk dilemma game provides an adequate analytical approach to capture this conflict. In these models, it is generally assumed that the risk of collective failure is perceived with certainty, but the real-world complexities are better captured by risk uncertainty. To address this shortage, we consider the collective-risk dilemma game with uncertain risk. In the proposed game, in addition to unconditional cooperators and defectors, a third type adaptive strategy is introduced. They bear an observation cost to infer the corresponding risk level and adopt their subsequent action based on this information: cooperate if the inferred risk level exceeds their tolerance thresholds and defect, otherwise. We find that natural selection favors the evolution of adaptive agents, especially under the conditions of low observation cost and moderate tolerance threshold. Furthermore, a relatively low observation cost and tolerance threshold result in a higher cooperation level. This observation remains intact if strategy mutations are allowed. In addition, under certain conditions, we find that the introduction of adaptive agents promotes cooperation and group achievement significantly compared to the two-strategy scenarios.
\end{quotation}

\section{Introduction}
Cooperation is essential for the survival and prosperity of many species, including humans~\cite{Axelrod1981Science,Rong2007PRE,Perc2017PR,Aktipis2018NHB,Jusup2022PR,Han2024JRSI,Han2025CSF,Wang2026PNAS,Morozov2026PNAS,Allen2026NCS,Li2026reinforcement}. This altruistic action, in which individuals entail personal costs to generate benefits for others, can be observed in various contexts, such as combating global warming~\cite{Jacquet2013NCC}, managing common resources~\cite{Ostrom1990}, and curbing the spread of infectious diseases~\cite{Bavel2020NHB}. However, the emergence and maintenance of cooperation pose a challenging enigma, stemming from the tension between individual and collective interests~\cite{Han2026PLF}. Evolutionary game theory offers a rigorous and widely adopted analytical framework for addressing this conundrum~\cite{Hofbauer1998,Nowak2006,Sigmund2010}. A variety of simple yet effective games, including the Prisoner's Dilemma~\cite{Nowak1992Nature,Liu2015EL,Wang2023JRSI}, the Snowdrift Game~\cite{Hauert2004Nature,Doebeli2004Science}, the Stag Hunt Game~\cite{Pacheco2009PRSB,Luo2021PD}, and the Public Goods Game~\cite{Perc2013JRSI,Han2024Chaos,Chen2024DGA,Hauert2024JRSI,Zhu2024CP,Wei2025IRSI,Wang2026SA,Jiang2026nonlinear,Wang2020steering}, have received considerable attention in both theoretical and experimental research. 

Of particular note is the collective-risk dilemma (CRD) game is regarded as a paradigm for investigating how cooperation evolves in the face of dangerous climate change or analogous problems~\cite{Milinski2006PNAS,Milinski2008PNAS,Tavoni2011PNAS,Quan2023Chaos}. A substantial body of empirical and theoretical research has demonstrated that the CRD game plays a crucial role in fostering the evolution of cooperative behavior~\cite{Pacheco2009PRSB,Santos2011PNAS,Vasconcelos2013NCC,Chen2012PRE,Chen2014PRE, He2019AMC}. For example,  Santos~{\it et al.} found that a higher risk of loss facilitates the evolution of cooperation, while a lower risk
triggers its collapse~\cite{Santos2011PNAS}. Vasconcelos~{\it et al.} later introduced sanctioning institutions to deter free riders. Their findings showed that the implementation of such institutions can enhance the levels of cooperation, and local institutions are more effective than global ones~\cite{Vasconcelos2013NCC}. Liu~{\it et al.} developed a coevolutionary game approach that can capture the coupled dynamics of cooperation and risk. The results revealed that cooperation can be maintained by keeping risk at a specific level or forming evolutionary oscillations with risk~\cite{Liu2023EL}. In addition, Hua~{\it et al.} proposed an evolutionary game model with repeated interactions and demonstrated that the introduction of a 
conditional cooperation strategy can effectively catalyze the evolution of cooperation, especially in low-risk scenarios~\cite{Hua2023PRSB}.

Although the existing literature has made significant strides in mitigating the cooperative dilemma, they often treat the risk level as a fixed value~\cite{Santos2012MMMAS,Vasconcelos2013NCC,Hua2023PD,Wang2026PLOSCB}. In reality, however, the risk in social environments is inherently uncertain and may influence individual's decision-making~\cite{szolnoki_pre15}. Notably, some individuals in a population may adaptively modulate their behavioral actions in response to such environmental uncertainty. This perspective has already been incorporated into theoretical frameworks investigating the evolution of cooperation~\cite{Colnaghi2023PNAS}. Building on this, we consider the CRD game with uncertain risk. In such a game, we introduce a sort of adaptive agent who incurs an observation cost to estimate the risk magnitude for each interaction. These adaptive agents then take  actions based on the corresponding tolerance thresholds: they will cooperate when the risk level is greater than the thresholds and defect, otherwise. When the CRD game includes three different strategies (adaptive agents, always cooperate, and always defect), two fundamental but unresolved questions therefore arise: (i) how do these adaptive agents evolve in the CRD game with uncertain risk? and (ii) how does their presence affect the cooperation level in the population?

In this paper, we therefore study the evolution of adaptive agents and their impact on the cooperation level in the CRD game with uncertain risk. To quantify risk uncertainty, we model the risk of collective disaster as a random variable following a uniform distribution. Our analysis first focuses on a finite and well-mixed population in the absence of mutation. In this case, we numerically calculate the fixation probabilities of the three types of agents for different observation costs and tolerance thresholds. Our results demonstrate that natural selection confers an evolutionary advantage to adaptive agents under conditions of low observation costs and moderate tolerance thresholds. Based on the corresponding fixation probabilities, we further compute the average cooperation level within the population for distinct combinations of observation costs and tolerance thresholds. The results show that small observation costs and tolerance thresholds facilitate higher cooperation levels in the presence of adaptive agents. Furthermore, we explore how the presence of adaptive agents affects the average cooperation level and group achievement in the population when 
mutation are considered. For this situation, we first determine the stationary distribution of system and then compute the selection gradient of population. We further assess the average cooperation level and group achievement across different observation costs, tolerance thresholds, and mutation rates, respectively. Notably, we observe that the trend of the average cooperation level with increasing observation cost or tolerance threshold is similar to that of the case without mutation. We further find that at low tolerance thresholds, both the average cooperation level and group achievement increase with the mutation rate. However, as the tolerance threshold rises, these two quantities first increase and then decrease as the mutation rate increases. Moreover, the findings suggest that introducing adaptive agents leads to a higher cooperation level and group achievement compared to the two-strategy cases. However, as the tolerance threshold increases, this advantage diminishes, and these two quantities are inferior to the levels observed in the two-strategy scenarios.

\section{Model and methods}
\subsection{Collective-risk social dilemma with uncertain risk}
\begin{figure*}
\centering
\includegraphics[width=5.0in]{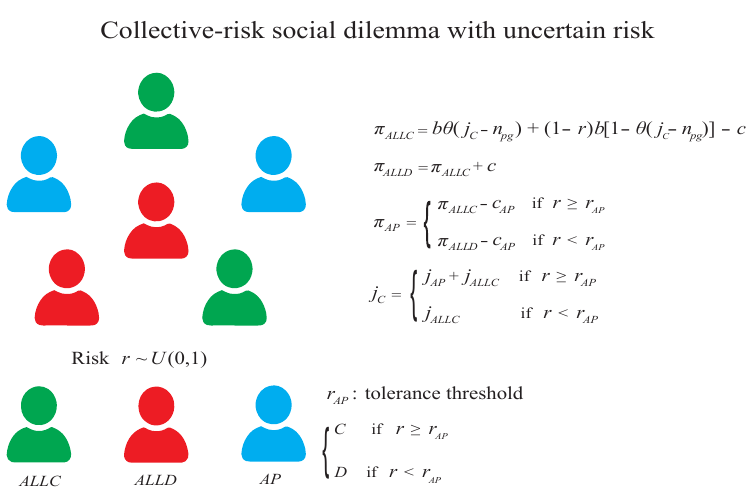}
\caption{Sketch of the collective-risk social dilemma with uncertain risk.  
In each round, $N$ agents participate in the collective-risk social dilemma game. These agents include adaptive agents ($AP$) and two kinds of agents who always cooperate ($ALLC$) or always defect ($ALLD$). The risk level in the game, denoted as, $r$, is drawn from a uniform distribution on $[0, 1]$.  Adaptive agents incur a cost to infer the risk level in each interaction. Their subsequent action depends on their tolerance thresholds, $r_{AP}$. To be specific, they cooperate if $r \geq r_{AP}$, and defect otherwise. Each agent has an initial endowment $b$. Cooperators contribute a fixed $c$ to achieve a common goal, while defectors contribute nothing. If the total number of $ALLCs$ and $APs$ who cooperate in the group is below a threshold, $n_{pg}$, every agent in that group will lose their remaining endowment with a probability $r$. Otherwise, everyone will keep whatever they have. When adaptive agents are present, the payoffs of an $ALLC$, $ALLD$, and $AP$ in a group including $j_C$ contributors are given by $\pi_{ALLC}$, $\pi_{ALLD}$, and $\pi_{AP}$, respectively.}
\label{fig1}
\end{figure*}

Consider a finite well-mixed population of size $Z$, from which $N$ agents are selected to engage in the CRD game. Each agent receives an initial endowment $b$ and adopts either cooperation ($C$) or defection ($D$). Cooperators contribute a fixed $c$ to achieve a common goal, whereas defectors contribute nothing. If the number of cooperators in the group falls below a threshold $n_{pg}$ ($0<n_{pg}< N$), every member of that group will lose their remaining endowment with probability $r$ ($0 \leq r \leq 1$). Otherwise, everyone will keep their benefits. Accordingly, the payoff of a cooperator playing in a group in which there are $j_C$ cooperators and $N-j_C$ defectors can be written as
\begin{equation}
\pi_C= b \theta(j_C-n_{pg})+(1-r)b[1-\theta(j_C-n_{pg})] -c,\label{Eq1}
\end{equation}
and the payoff of a defector in a group with $j_C$ cooperators and $N-j_C$ defectors is given by
\begin{equation}
\pi_D=b \theta(j_C-n_{pg})+(1-r)b[1-\theta(j_C-n_{pg})].\label{Eq2}
\end{equation}
Here, $\theta(x)$ denotes the Heaviside function, defined as $\theta(x)=1$ for $x\ge 0$ and $\theta(x)=0$ for $x <0$. Unlike traditional CRD games, we investigate the collective-risk social dilemma game with a risk uncertainty. To better enhance the intuitive understanding of this game, a schematic illustration is presented in Fig.~\ref{fig1}. We assume that the risk level in the game follows a uniform distribution, given as $r \sim U(0, 1)$. We emphasize that the uniform distribution is adopted as our baseline choice because it well captures the uncertainty of risk, $r$, in the absence of external influences. Furthermore, we introduce a kind of adaptive agents ($AP$), whose decision to cooperate or defect is contingent upon the risk level in each interaction. Specifically, when interacting with two agents, who either always cooperate ($ALLC$) or always defect ($ALLD$), within an $N$-person group, these adaptive agents pay a cost $c_{AP}$, to infer the risk level.  After inferring the risk level, adaptive agents can take different actions based on the inferred information: they will cooperate if the risk level is higher than their own tolerance threshold $r_{AP} $, otherwise, they will defect. Thus, based on Eqs.~\eqref{Eq1} and~\eqref{Eq2}, we calculate the payoffs of the three different strategists in a single game as follows:
\begin{equation}
\pi _{ALLC}=b \theta(j_C-n_{pg})+(1-r)b[1-\theta(j_C-n_{pg})] -c, 
\end{equation}
\begin{equation}
\pi _{ALLD}=\pi _{ALLC}+c ,
\end{equation}
\begin{equation}
\pi _{AP}=\left\{\begin{matrix}
 \pi _{ALLC} -c_{AP} \;\; \text{if} \;\; r\ge r_{AP}, \\\pi _{ALLD} -c_{AP}\;\;\text{if} \ r <  r_{AP}.
\end{matrix}\right.
\end{equation}
Here, $j_{C}$ represents the total number of $ALLC$ and $APs$ who cooperate in the group.

\subsection{Strategy update}
We then consider the strategy update process of the population. At each time step, an $X$-agent is randomly selected to update its strategy. With probability $1-u$, this process follows the commonly used pairwise comparison (PC) rule~\cite{Ohtsuki2006JTB,Han2025CSF,Wang2026SA}. To be specific, an agent with $X$-strategy is randomly chosen for updating its strategy. Then it chooses a $Y$-agent at random. The probability that this $X$-agent adopts the strategy of $Y$-agent is given by~\cite{Szabo1998PRE}
\begin{equation}
f=\frac{1}{1+e^{-w(f_Y- f_X)} },
\end{equation}
where $f_X$ and $f_Y$ denote the payoffs of agents $X$ and $Y$. The non-negative parameter $w$ measures the intensity of selection. In this work, we focus on moderately strong selection by setting $w=0.5$. In addition, we introduce the behavioral mutation, where with probability $u$, $X$-agent undergoes a mutation, adopting a strategy drawn randomly from the space of available strategies. Therefore, in a three-strategy population, the transition probability for an individual currently adopting strategy
$X$ to switch to a different strategy $Y$ under the PC rule with mutation rate $u$, given as
\begin{equation}
T_{X \rightarrow Y}=(1-u) \frac{i_{X}}{Z} \frac{i_{Y}}{Z-1}\frac{1}{1+e^{-w\left(f_{Y}-f_{X}\right)}}+u \frac{i_{X}}{ 2Z}, \label{Eq10}
\end{equation}
where $i_{X}$ and $i_{Y}$ are the numbers of $X$ and $Y$ agents in the population. Therefore, for a given configuration $\mathbf{i}=\left ( i_{1},i_{2},i_{3} \right )$, the probabilities that the number of adaptive agents increases or decreases by one, denoted by $T_{\mathbf{i}}^{AP+}$ and $T_{\mathbf{i}}^{AP-}$, are expressed as
\begin{equation}
T_{\mathbf{i}}^{AP \pm}=T_{\left(i_{1} \pm 1, i_{2} \mp 1, i_{3}\right)}+T_{\left(i_{1} \pm 1, i_{2}, i_{3} \mp 1\right)}.
\end{equation}
Similarly, the corresponding probabilities of $ALLC$ and $ALLD$ are, respectively,
\begin{equation}
T_{\mathbf{i}}^{ALLC \pm}=T_{\left(i_{1} \mp 1, i_{2} \pm 1, i_{3}\right)}+T_{\left(i_{1} , i_{2}\pm 1, i_{3} \mp 1\right)},
\end{equation}
and
\begin{equation}
T_{\mathbf{i}}^{ALLD \pm}=T_{\left(i_{1} \mp 1, i_{2}, i_{3}\pm 1\right)}+T_{\left(i_{1} , i_{2}\mp 1, i_{3} \pm 1\right)}.
\end{equation}

Based on the formulas above, we can compute the selection gradient to characterize the most likely evolutionary path of the population as it leaves the configuration $\mathbf{i}$. 
The
configurations are parameterized as $\mathbf{i}=( i_{1},i_{2}  )$, where $i_1$ and $i_2$ are the counts of $APs$, $ALLCs$ (and $Z-i_1-i_2$ $ALLDs$), respectively. The selection gradient for this three-strategy system is thus expressed as
\begin{equation}
\nabla_{i}=\left(T_{i}^{AP+}-T_{i}^{AP-}\right) u_{AP}+\left(T_{i}^{ALLC+}-T_{i}^{ALLC-}\right) u_{ALLC},\label{Eq14}
\end{equation}
where $u_{AP}$ and $u_{ALLC}$ are basis vectors.

\subsection{Average payoff calculations for individuals with  different strategies}\label{SubSec:III}
Based on the foregoing description, we proceed to calculate the average payoffs of three distinct strategists. We assume that each agent engages in a finite number of game rounds prior to strategy updating, and whether it revises its strategy depends on the average payoff obtained from these game interactions. To achieve this, we perform $m$ subsamplings of the risk level of collective failure, denoted as $\mathbf{r}=\left ( r_{1} ,r_{2}, \cdots ,r_{m}  \right) $. Using these $m$ sampled risk values, we then determine the individual average payoff for each type of agents. Specifically, the payoff of an agent playing in a group where there are $j_{1}$ $APs$, $j_{2}$ $ALLCs$ and $N-j_{1}-j_{2}$ $ALLDs$ is respectively expressed as
\begin{align}
\begin{split}
\Pi_{AP}(j_{1},j_{2})= & -c_{AP}\\
&+\frac{1}{m} \sum_{r_{i} \geq r_{AP}}\Big\{-c+b \theta\left(j_{1}+j_{2}-n_{p g}\right)\\
&
\hspace{1.7cm}
+\left(1-r_{i}\right) b\big[1-\theta\left(j_{1}+j_{2}-n_{p g}\right)\big]\Big\} \\
& +\frac{1}{m} \sum_{r_{i} < r_{AP}}\Big\{b \theta\left(j_{2}-n_{p g}\right)\\
& 
\hspace{1.7cm}
+\left(1-r_{i}\right) b\big[1-\theta\left(j_{2}-n_{p g}\right)\big]\Big\},
\end{split}
\end{align}
\begin{align}
\begin{split}
\Pi_{ALLC}(j_{1},j_{2})=&\frac{1}{m} \sum_{r_{i} \geq r_{AP}}\Big\{-c+b \theta\left(j_{1}+j_{2}-n_{p g}\right)\\
& 
\hspace{1.4cm}
+\left(1-r_{i}\right) b\big[1-\theta\left(j_{1}+j_{2}-n_{p g}\right)\big]\Big\} \\
& +\frac{1}{m} \sum_{r_{i} < r_{AP}}\Big\{-c+b \theta\left(j_{2}-n_{p g}\right)\\
& 
\hspace{1.4cm}
+\left(1-r_{i}\right) b\big[1-\theta\left(j_{2}-n_{p g}\right)\big]\Big\},
\end{split}
\end{align}
and
\begin{align}
\begin{split}
\Pi_{ALLD}(j_{1},j_{2}) = &\frac{1}{m} \sum_{r_{i} \geq r_{AP}}\Big\{b \theta\left(j_{1}+j_{2}-n_{p g}\right)\\
& 
\hspace{1.4cm}
+\left(1-r_{i}\right) b\big[1-\theta\left(j_{1}+j_{2}-n_{p g}\right)\big]\Big\} \\
& +\frac{1}{m} \sum_{r_{i} < r_{AP}}\Big\{b \theta\left(j_{2}-n_{p g}\right)\\
& 
\hspace{1.7cm}
+\left(1-r_{i}\right) b\big[1-\theta\left(j_{2}-n_{p g}\right)\big]\Big\}.
\end{split}
\end{align}

Subsequently, we calculate the final average payoffs of these three agents. We assume that there are $i_1$ $APs$, $i_2$ $ALLCs$, and $i_3$ $ALLDs$ in a population of size $Z$. Let $j_1$, $j_2$, and $j_3$ denote the number of $APs$, $ALLCs$, and $ALLDs$ in a game group of size $N$, respectively. The final average payoff of $AP$, $ALLC$, and $ALLD$ can be expressed as
\begin{equation}
f_{AP}=\sum_{j_{2}=0}^{N-1} \sum_{j_{1}=0}^{N-1-j_{2}} \frac{\binom{i_{1}-1}{j_{1}}\binom{i_{2}}{j_{2}}\binom{Z-i_{1}-i_{2}}{N-1-j_{1}-j_{2}}}{\binom{Z-1}{N-1}} \Pi_{AP}\left(j_{1}+1, j_{2}\right),
\end{equation}
\begin{equation}
f_{ALLC}=\sum_{j_{2}=0}^{N-1} \sum_{j_{1}=0}^{N-1-j_{2}} \frac{\binom{i_{1}}{j_{1}}\binom{i_{2}-1}{j_{2}}\binom{Z-i_{1}-i_{2}}{N-1-j_{1}-j_{2}}}{\binom{Z-1}{N-1}} \Pi_{ALLC}\left(j_{1}, j_{2}+1\right),
\end{equation}
and
\begin{equation}
f_{ALLD}=\sum_{j_{2}=0}^{N-1} \sum_{j_{1}=0}^{N-1-j_{2}} \frac{\binom{i_{1}}{j_{1}}\binom{i_{2}}{j_{2}}\binom{Z-i_{1}-i_{2}-1}{N-1-j_{2}-j_{1}}}{\binom{Z-1}{N-1}} \Pi_{ALLD}\left(j_{1}, j_{2}\right).
\end{equation}
Here, the binomial coefficient $\binom{i}{j}=0$ when $i<j$.

\subsection{Evolutionary dynamics without and with behavioral mutation}
We model the evolutionary dynamics of a population where two types of agents that always cooperate ($ALLC$) or always defect ($ALLD$) compete with adaptive agents ($AP$). Each possible population state is defined by a triplet ($i_1$, $i_2$, $i_3$), where $i_1$, $i_2$, and $i_3$ represent the number of adaptive agents, $ALLC$, and $ALLD$, respectively. In the initial state, we assume that the three strategies are uniformly distributed within the population. The evolution of the population is described as a stochastic process governed by transition probabilities. Using Eq.~\eqref{Eq10}, we can derive the transition matrix $T$, whose elements $T_{ij}$ denote the transition probabilities from state $i$ to state $j$. 

We first investigate the case of $u=0$, in which the system has three absorbing states: $AP$ fixation $(Z, 0, 0)$, $ALLC$ fixation $(0, Z, 0)$, and $ALLD$ fixation $(0, 0, Z)$. To determine the probabilities of reaching these three states from any initial transient state, we re-express the transition matrix $T$ in the following canonical form
\begin{equation}
P=\left[\begin{array}{ll}
Q & R \\
0 & I_3
\end{array}\right].\label{Eq15}
\end{equation}
Here, $Q$ is a $t \times t$ matrix whose entries represent the transition probability between transient states, where $t$ is the number of transient states, $R$ is a $t \times 3$
matrix that denotes the transition probabilities from each transient state to each absorbing state, and $I_3$ is a $3 \times 3$ identity matrix~\cite{Grinstead1997}. The probability 
of reaching an ergodic state $j$ starting from a transient state $i$ is given by the $(i,j)$ -- the entry $b_{ij}$ of the matrix $B$, given as
\begin{equation}
B=\left (I_{t}-Q \right ) ^{-1} R.  \label{Eq16}
\end{equation}
Using Eq.~\eqref{Eq16}, we can compute the probabilities of reaching the absorbing states $(Z,0,0)$, $(0,Z,0)$, and $(0,0,Z)$ starting from the uniform state $(Z/3, Z/3, Z/3)$.  

Let $\phi_{A}$ and $\phi_C$ denote the fixation probability of $AP$ and $AllC$, respectively. If adaptive agents cooperate with a frequency $\alpha$, a parameter jointly determined by the tolerance threshold and the distribution of the risk level, then the average cooperation level is given by~\cite{Colnaghi2023PNAS} 
\begin{equation}
P_{C} = \phi _{C}  + \alpha \phi_{A} .\label{Eq17}
\end{equation}

Subsequently, we consider the case where mutation probability is non-zero. In this regime, the system has a stationary distribution, $\bar{p}$, representing the amount of time that the population spends at each state. This distribution can be analytically computed by normalizing the eigenvector associated with the eigenvalue $1$ of the transition matrix $T$ of the Markov chain. In this case, we define the average cooperation level in the population as
\begin{equation}
P_{C} =\frac{1}{Z} \sum_{\mathbf{i}=\left ( i_{1},i_{2} ,i_{3}  \right ) }^{} \bar{p_{\mathbf{i}}} \left ( \alpha i _{1}+i_{2}  \right ) .\label{Eq18}
\end{equation}
Furthermore, for each possible configuration $\mathbf{i}$, we use multivariate hypergeometric sampling to calculate the fraction of groups that reach $n_{pg}$ contributors, given as
\begin{equation}
a_{G}\left ( \mathbf{i} \right )=\sum_{j_{2}=0 }^{N} \sum_{j_{1} =0}^{N-j_{2} }\frac{ \binom{i_1}{j_1}\binom{i_2}{j_2}\binom{Z-i_1-i_2}{N-j_1-j_2}}{\binom{Z}{N}}g(j_1,j_2),
 \end{equation}
where $g(j_1,j_2)= \alpha \theta \left ( j_{1}+j_{2}-n_{pg} \right )+\left ( 1-\alpha  \right ) \theta \left ( j_{2}-n_{pg} \right ) $. We then compute another important quantity, the average group achievement, $\eta_G$, by averaging over all possible configurations $\mathbf{i}$, weighted by the corresponding stationary distribution~\cite{Vasconcelos2013NCC}
\begin{equation}
\eta _{G} = \sum_{\mathbf{i}=\left ( i_{1},i_{2} ,i_{3}  \right ) }^{} \bar{p_{\mathbf{i}}}a_{G}\left ( \mathbf{i} \right )  .\label{Eq20}
\end{equation}

\section{Results}
\subsection{Evolutionary outcomes without mutations}
First, we investigate the evolutionary dynamics in the absence of mutation. The population state is represented by a triplet $(i_1, i_2, i_3)$, where $i_1$, $i_2$, and $i_3$ denote the number of $AP$, $ALLC$, and $ALLD$, respectively. When $u=0$, the system exhibits three absorbing states: $AP$ fixation $(Z, 0, 0)$, $ALLC$ fixation $(0, Z, 0)$, and $ALLD$ fixation $(0, 0, Z)$. Using the formalism described in Eq.~\eqref{Eq16}, we calculate the probability of reaching these three states starting from an initial state where all strategies are present in equal proportion $( Z/3,  Z/3, Z/3 )$. Furthermore, according to Eq.~\eqref{Eq17}, we evaluate the average  cooperation level in the population.

Fig.~\ref{fig2} illustrates how the observation cost, $c_{AP}$, affects the fixation probabilities of $AP$, $ALLC$, and $ALLD$ for different tolerance thresholds, $r_{AP}$. Across all tested $r_{AP}$ values, the fixation probability of $AP$ declines monotonically as $c_{AP}$ increases, as shown in Fig.~\ref{fig2}(a). By contrast, the fixation probabilities of both $ALLC$ and $ALLD$ exhibit an increasing trend with rising $c_{AP}$, respectively, as shown in Fig.~\ref{fig2}{(b) and (c). Furthermore, we find that natural selection more readily favors the evolution of adaptive agents when observation costs are low and tolerance thresholds are moderate, resulting in fixation probabilities higher than the neutral expectation. However, as the observation cost increases or the tolerance threshold becomes too small or too large, this evolutionary advantage gradually erodes and vanishes.
\begin{figure*}
\centering
\includegraphics[width=\textwidth]{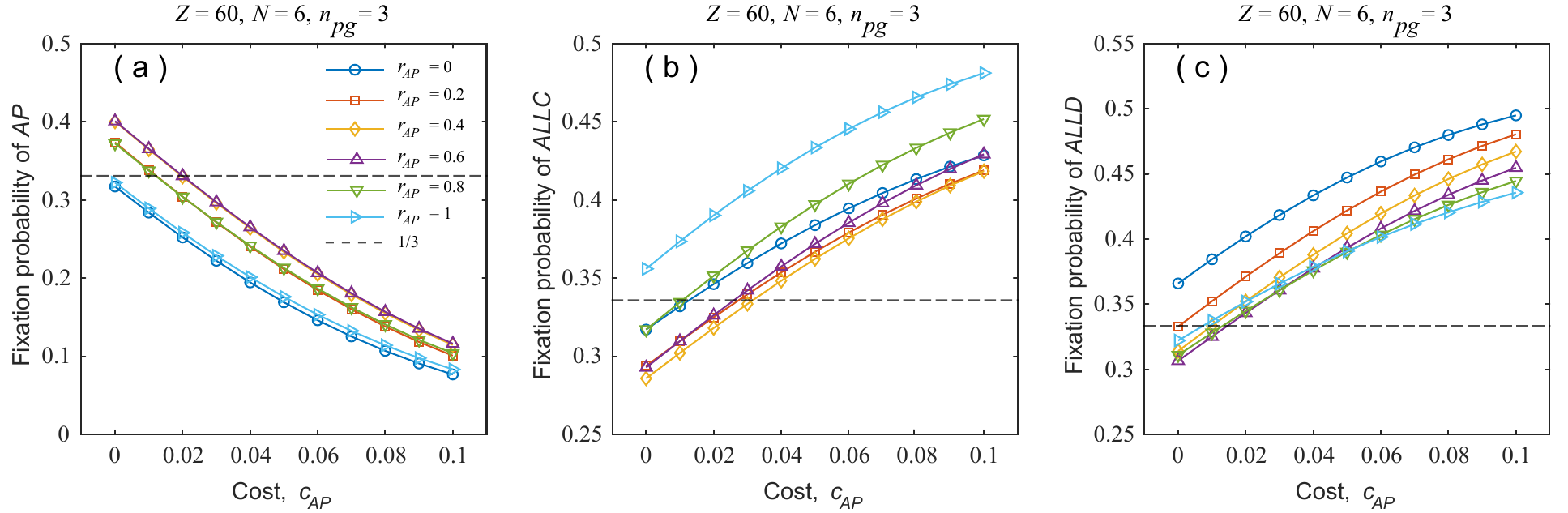}
\caption{Fixation probabilities of three types of agents as a function of the observation cost $c_{AP}$ for different values of $r_{AP}$. In panel~(a), it shows that as $c_{AP}$ increases, $AP$ fixation probability decreases monotonically. In panels~(b) and (c), the fixation probabilities of $ALLC$ and $ALLD$ exhibit a monotonically increasing trend with rising $c_{AP}$, respectively. The horizontal dashed line represents the neutral fixation probability, that is, $P = 1/3$. Parameters: $Z = 60$, $N = 6$, $c = 0.1$, $b = 1$, $n_{pg} = 3$.
}
\label{fig2}
\end{figure*}

Next, we assess how the fixation probabilities of $AP$, $ALLC$, and $ALLD$ are modulated by the tolerance threshold, $r_{AP}$, across different observation costs, $c_{AP}$. Fig.~\ref{fig3}(a) shows that across all tested $c_{AP}$ values, the fixation probability of $AP$ initially increases and then decreases with respect to $r_{AP}$, peaking at $r_{AP}\approx0.5$. In contrast, as depicted in Fig.~\ref{fig3}(b), $ALLC$ fixation probability first drops then rises with increasing $r_{AP}$, reaching its minimum at $r_{AP} \approx 0.4$. Fig.~\ref{fig3}(c) indicates that the fixation probability of $ALLD$ first decreases then increases at small $c_{AP}$, whereas it declines monotonically at larger $c_{AP}$ values. Moreover, we find that, for most parameter combinations of $c_{AP}$ and $r_{AP}$, adaptive agents fixate with a probability below the neutral baseline of $1/3$. These findings further imply that a lower observation cost coupled with a moderate tolerance threshold best facilitates the evolutionary success of adaptive agents.
\begin{figure*}
\centering
\includegraphics[width=\textwidth]{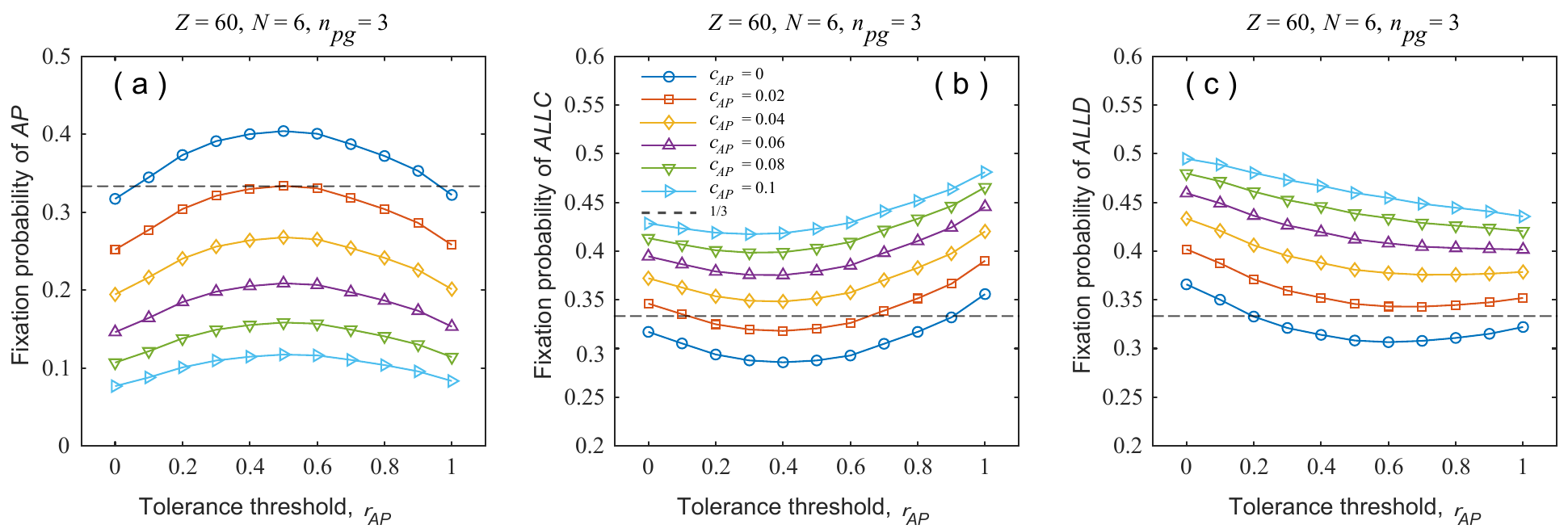}
\caption{Fixation probabilities of three different agents in dependence on the tolerance threshold $r_{AP}$ under different values of $c_{AP}$. In panel~(a), it shows that $AP$ fixation probability increases first and then decreases as $r_{AP}$ increases. In panel~(b), it shows that when $r_{AP}$ increases, the fixation probability of $ALLC$ first decreases and then rises. In panel~(c), when $c_{AP}$ is small, the fixation probability of $ALLD$ decreases first and then increases, while it monotonically decreases in other cases. The horizontal dashed line represents the neutral fixation probability, namely, $P = 1/3$. Parameters: $Z = 60$, $N = 6$, $c = 0.1$, $b = 1$, $n_{pg} = 3$.}
\label{fig3}
\end{figure*}

We further examine how $c_{AP}$ and $r_{AP}$ influence average cooperation level in a population where adaptive agents, cooperators, and defectors are equally distributed. In Fig.~\ref{fig4}(a), it is observed that when $r_{AP}<0.5$, the average cooperation level decreases as $c_{AP}$ increases, whereas it increases as  $c_{AP}$ rises for $r_{AP}>0.5$. This outcome stems from the different action preferences of adaptive agents for distinct tolerance thresholds. When $r_{AP} <0.5$, adaptive agents tend to favor cooperation. However, increasing the observation cost for these agents reduces their fixation probability, which in turn hinders the evolution and maintenance of cooperation in the population. In contrast, when $r_{AP}>0.5$, adaptive agents tend to favor defection. In this regime, raising their observation cost lowers their fixation probability but enhances that of $ALLC$, thereby increasing the average cooperation level in the population. To assess the impact of adaptive agents on cooperation level, we also compute the average cooperation level in the two-strategy scenario with only $ALLC$ and $ALLD$. It can be found that for smaller $r_{AP}$ and $c_{AP}$, the introduction of adaptive agents helps to improve the group cooperation level. In Fig.~\ref{fig4}(b), it can be found that for many different value of $c_{AP}$, the average cooperation level declines with the increase of the tolerance threshold $r_{AP}$. This occurs because a higher tolerance threshold makes adaptive agents more likely to defect, thereby hindering group cooperation. When $r_{AP}$ is sufficiently low, adaptive agents can enhance group cooperation compared with the two-strategy scenario. Moreover, larger value of $ c_{AP}$ decreases the slopes of the curves, indicating that higher observation cost renders the cooperation level less responsive to changes in $r_{AP}$.
\begin{figure*}
\centering
\includegraphics[width=6.5in]{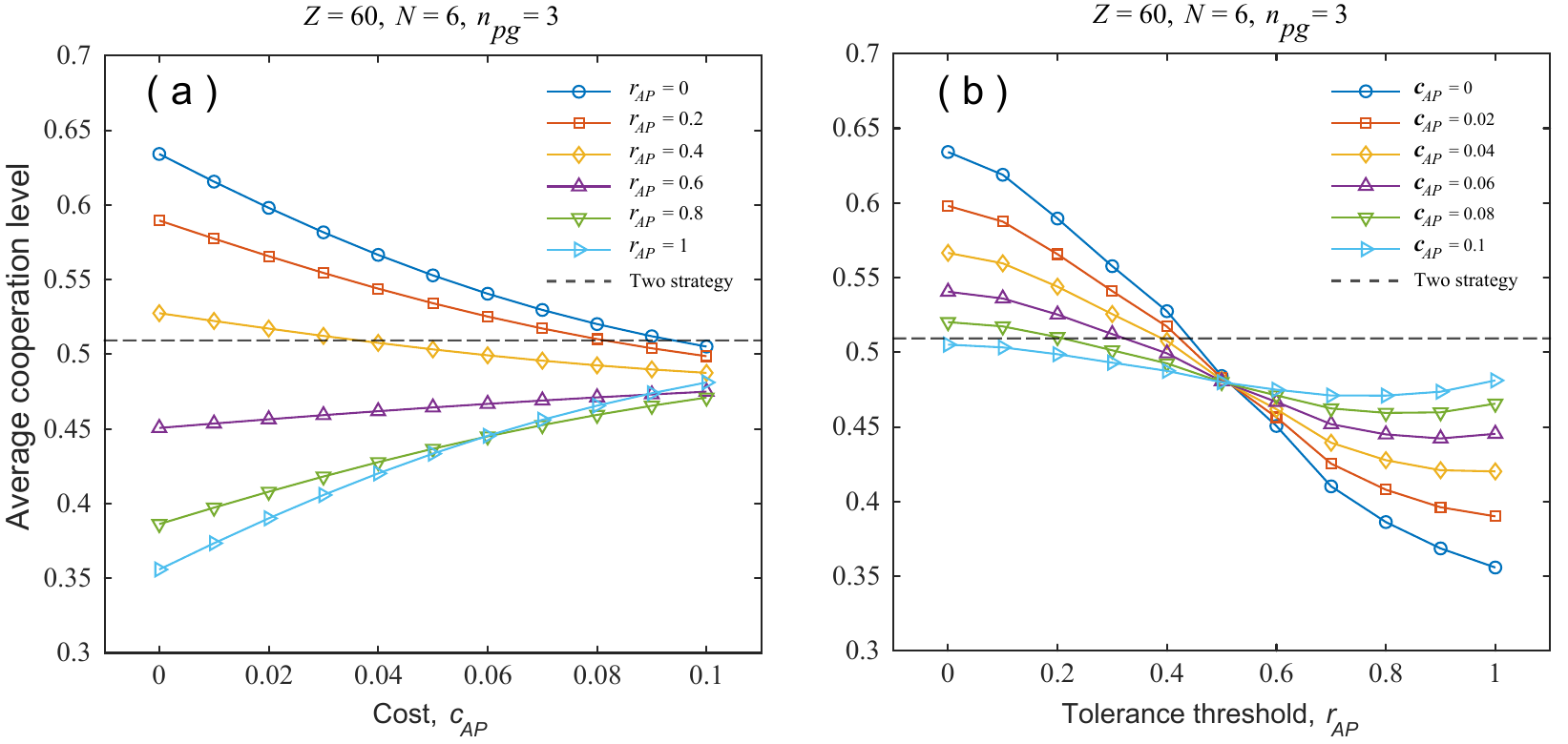}
\caption{In panel~(a), average cooperation level as a function of $c_{AP}$ under distinct values of $r_{AP}$. It shows that for small $r_{AP}$, the cooperation level decreases monotonically with the rise of $c_{AP}$, whereas it increases monotonically under large $r_{AP}$.  In panel~(b), average cooperation level as a function of $r_{AP}$ . In most cases, the average cooperation level decreases monotonically with the increase of $r_{AP}$. The black dashed line denotes the average cooperation level in the two-strategy scenario with only $ALLC$ and $ALLD$. Parameters: $Z = 60$, $N = 6$, $c = 0.1$, $b = 1$, $n_{pg} = 3$.}
\label{fig4}
\end{figure*}

\subsection{Evolutionary outcomes with mutations}
In this subsection, we study the evolutionary dynamics in the presence of mutations. The resulting system is governed by an ergodic Markov chain, which ensures the existence of a unique stationary distribution that characterizes the long-run fraction of time the population spends in each state. This distribution is derived by normalizing the left eigenvector corresponding to the unit eigenvalue of the state transition matrix, after which the selection gradient is computed using Eq.~\eqref{Eq14}.  We now present the stationary distribution and the selection gradient of the population under different combinations of parameters. In Figs.~\ref{fig5}(a-c), it can be found that when $r_{AP}$ is fixed but $c_{AP}$ is increased, the proportion of $AP$ strategy in the population gradually decreases. Furthermore, the shaded region shifts toward the $ALLC-ALLD$ boundary closer to the $ALLC$ vertex, indicating that the population tends to a state in which the $ALLC$ strategy will prevail. Similar trends are also observed for $r_{AP}=0.6$ and $r_{AP}=0.9$. In contrast, holding $c_{AP}$ constant while increasing $r_{AP}$ also reduces the fraction of adaptive agents in the population, yet it exerts a less pronounced effect on the $ALLD$ strategy. The results demonstrate that increasing either $c_{AP}$ or $r_{AP}$ diminishes the proportion of adaptive agents in the population. 
\begin{figure*}
\centering
\includegraphics[width=6.5in]{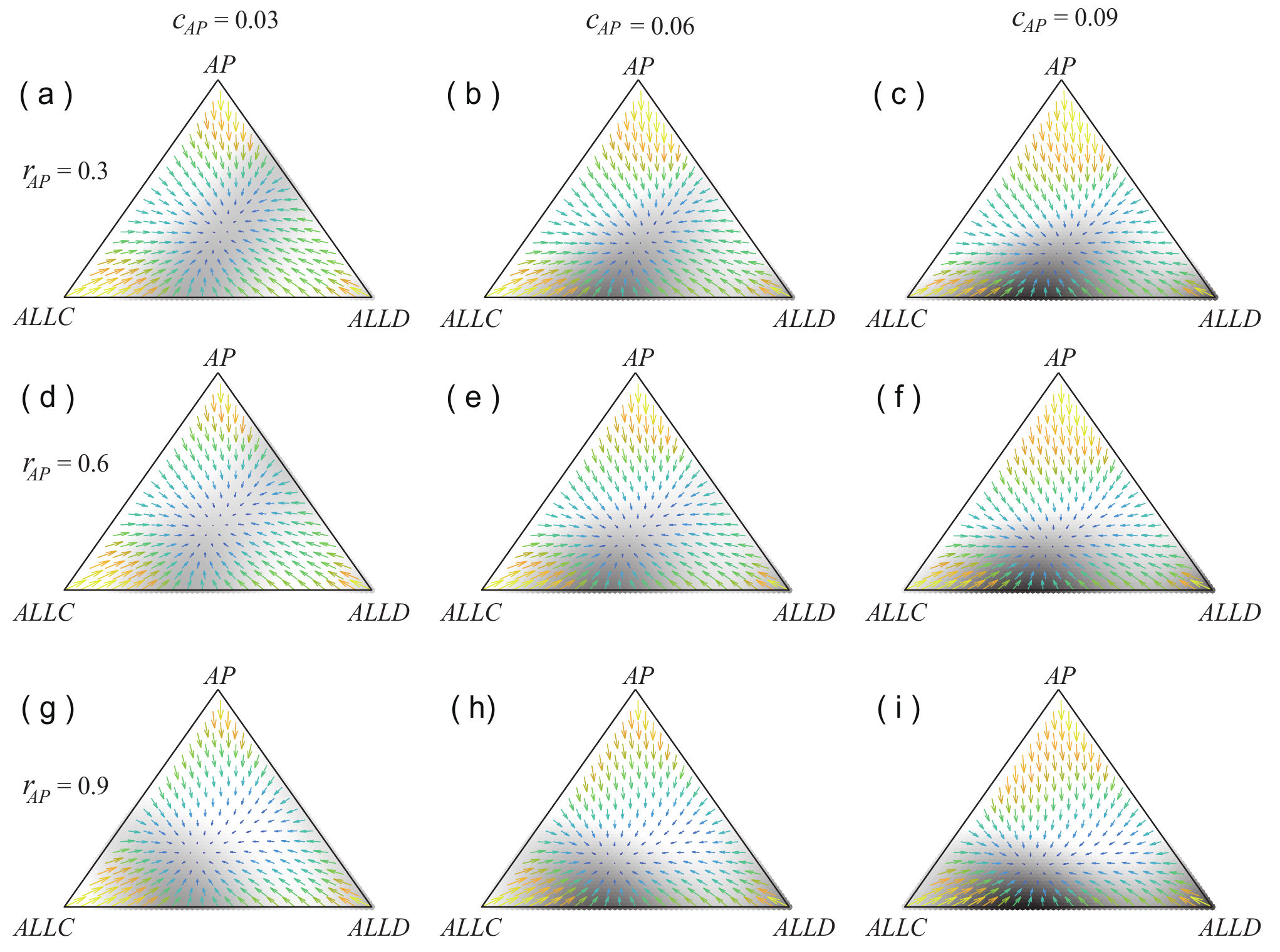}
\caption{Evolutionary dynamics  in finite populations with three different strategists for different values of $c_{AP}$ and $r_{AP}$. Each dot represents a state of the population. The color depth of the dots indicates the value of the stationary distribution for the corresponding state, with darker colors signifying states that are visited more frequently. The arrows denote selection gradients, which illustrate the most probable evolutionary direction of each state. The color intensity (and length) of the arrows reflects the magnitude of the gradient values. When $c_{AP}$ and $r_{AP}$ are small, the population remains for a long time in a state where $ALLD$ accounts for approximately half of the population. As $c_{AP}$ increases, the proportion of adaptive agents gradually declines, $ALLC$ quickly dominates the entire population, and the share of $ALLD$ also rises. A rise in $r_{AP}$ likewise reduces the fraction of adaptive agents, yet it exerts a less pronounced effect on the other two strategies compared with $c_{AP}$. Parameters: $Z = 60$, $N = 6$, $c = 0.1$, $b = 1$, $n_{pg} = 3$, $u=1/Z$.}
\label{fig5}
\end{figure*}

The triangular phase portrait alone cannot fully capture the population-level cooperation. We therefore show the two quantities, average cooperation level and group achievement, to further quantify the evolutionary outcomes in the presence of mutation.  These quantities are computed by using Eq.~\eqref{Eq18} and Eq.~\eqref{Eq20}, respectively. We first present how variations in $c_{AP}$ affect the average cooperation level in the population under different values of $r_{AP}$. As depicted in Fig.~\ref{fig6}(a), when $r_{AP}$ is smaller, the average cooperation level decreases as $c_{AP}$ rises, whereas it increases with $c_{AP}$ when $r_{AP}$ is larger. This phenomenon is attributed to the different action preferences of adaptive agents under different $r_{AP}$ values, and the specific explanation is similar to that in the case without mutation. In Fig.~\ref{fig6}(b), we show the group achievement of the population as a function of $c_{AP}$ for various values of $r_{AP}$. The results indicate that the group achievement decreases with increasing $c_{AP}$ at small $r_{AP}$, but increases at large $r_{AP}$. This trend closely parallels the variations in the average cooperation level observed in Fig.~\ref{fig6}(a). Furthermore, compared to
the two-strategy scenario involving only $ALLC$ and $ALLD$, the introduction of adaptive agents enhances cooperative behavior when $r_{AP}$ is small.
\begin{figure*}
\centering
\includegraphics[width=6.5in]{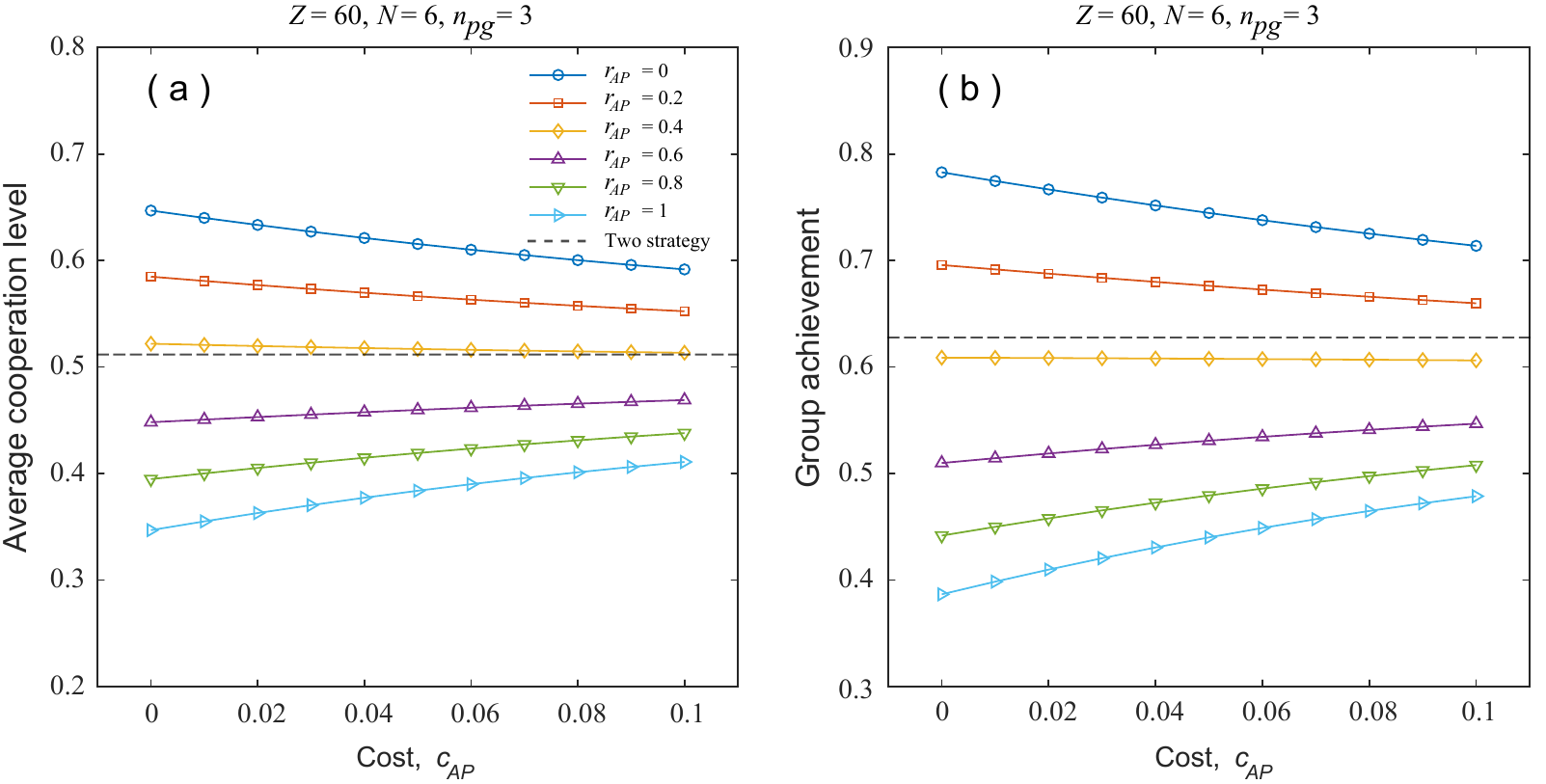}
\caption{In panel~(a), average cooperation level as a function of $c_{AP}$ for different values of $r_{AP}$. It shows that for small $r_{AP}$, the average cooperation level decreases monotonically with the increase of $c_{AP}$, whereas it increases monotonically under large $r_{AP}$. In panel~(b), group achievement as a function of $c_{AP}$ under different $r_{AP}$ . When $r_{AP}$ is small, group achievement decreases monotonically with the rise of $c_{AP}$, while it increases monotonically for large $r_{AP}$. The black dashed line represents the average cooperation level or group achievement in the two-strategy case, wherein the agents always cooperate or always defect. Parameters: $Z = 60$, $N = 6$, $c = 0.1$, $b = 1$, $n_{pg} = 3$, $u=1/Z$.}
\label{fig6}
\end{figure*}

We now present the average cooperation level and group achievement as functions of $r_{AP}$ for different values of $c_{AP}$, respectively. We find that the average cooperation level declines monotonically as $r_{AP}$ increases, as illustrated in  Fig.~\ref{fig7}(a). This occurs because a higher $r_{AP}$ encourages adaptive agents to choose defection, which undermines group cooperation and diminishes the average cooperation level. Therefore, a smaller $r_{AP}$ may be more conducive to fostering cooperation in the presence of adaptive agents. Furthermore, group achievement also deteriorates with increasing $r_{AP}$, as shown in Fig.~\ref{fig7}(b). This indicates that only smaller $r_{AP}$ can achieve higher group achievement.
\begin{figure*}
\centering
\includegraphics[width=6.5in]{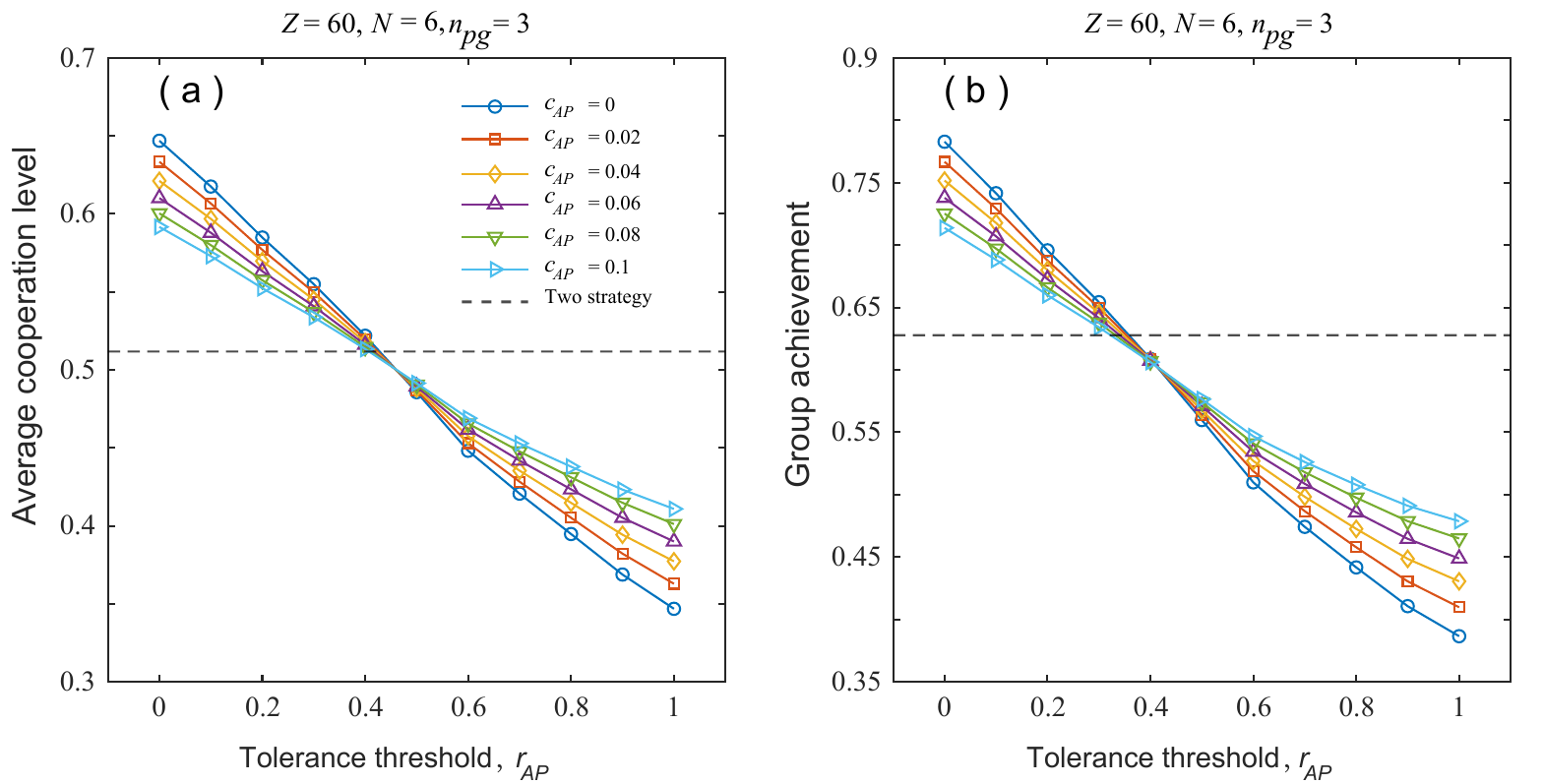}
\caption{In panel~(a), average cooperation level as a function of $r_{AP}$ under different values of $c_{AP}$. The average cooperation level decreases monotonically with the increase of $r_{AP}$. In panel~(b), group achievement as a function of $r_{AP}$, and the trend is aligned with that of the average cooperation level. The black dashed line denotes the average cooperation level in the two-strategy scenario in which each agent always cooperates or always defects. Parameters: $Z = 60$, $N = 6$, $c = 0.1$, $b = 1$, $n_{pg} = 3$, $u=1/Z$.}
\label{fig7}
\end{figure*}

Beyond the observation cost and tolerance threshold, the mutation rate $u$ is a critical determinant in the evolution of cooperation. Accordingly, we investigate how mutation modulates evolutionary dynamics in the presence of adaptive agents. To this end, we present the average cooperation level and group achievement as functions of $u$ for different values of $c_{AP}$ and $r_{AP}$. In Fig.~\ref{fig8}(a) and Fig.~\ref{fig9}(a), it can be observed that for small values of $r_{AP}$, increasing the mutation probability $u$ significantly enhances both average cooperation level and group achievement. However, as $r_{AP}$ increases, both quantities first rise and then fall with increasing $u$. Moreover, as the mutation rate increases, the average cooperation level and group achievement depend on the tolerance threshold, $r_{AP}$. Notably, when $r_{AP}$ is small, smaller values of $c_{AP}$ correspond to higher average cooperation level and group achievement, as depicted in Fig.~\ref{fig8}(a) and Fig.~\ref{fig9}(a), respectively. Conversely, when $r_{AP}$ is large, larger values of $c_{AP}$ lead to higher average cooperation level and group achievement, as shown in Fig.~\ref{fig8}(b-c) and Fig.~\ref{fig9}(b-c). These findings align with the conclusions drawn for scenarios without mutation or with low mutation rates. Therefore, our main finding remains robust to the introduction of mutation. Besides, compared to the two-strategy cases, the introduction of adaptive agents improves the average cooperation level and group achievement when tolerance thresholds are small. Notably, these two quantities are most effectively supported under the joint conditions of low observation costs and tolerance thresholds. Nevertheless, as the tolerance threshold increases, both the average cooperation level and the group achievement decreases in the population with adaptive agents, compared to the two-strategy case. 
\begin{figure*}
\centering
\includegraphics[width=\textwidth]{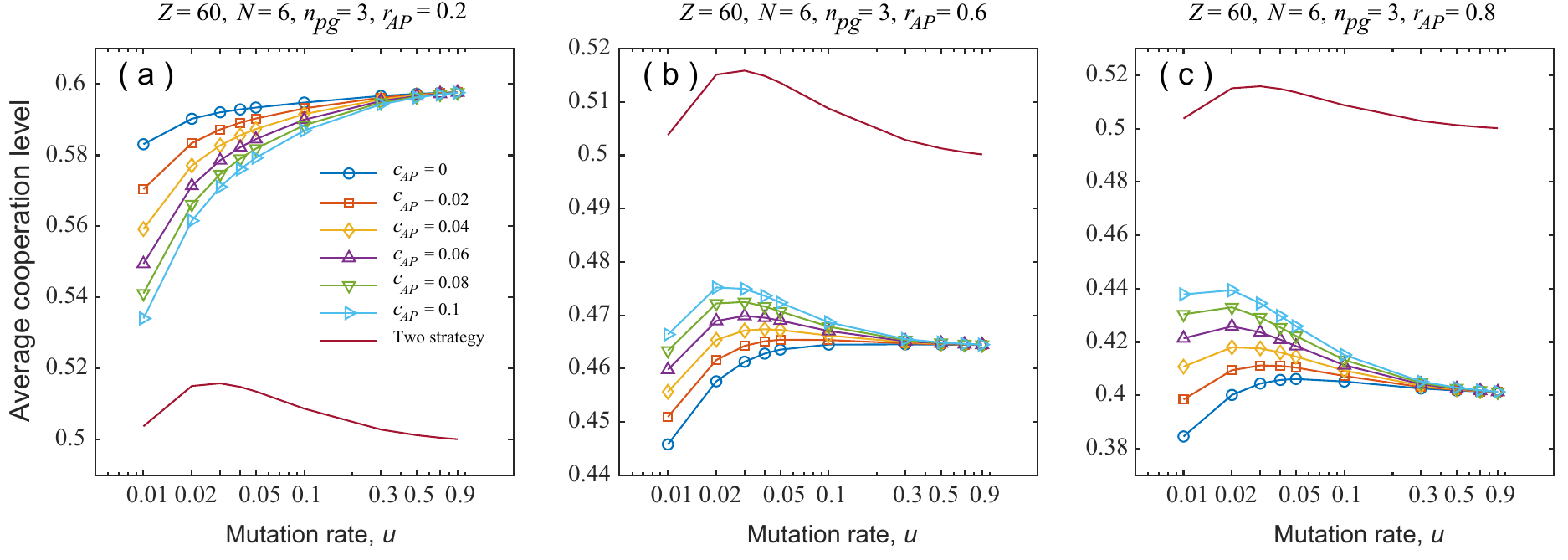}
\caption{Average cooperation level as a function of mutation rate $u$ for different combinations of $c_{AP}$ and $r_{AP}$. In panel~(a), where $r_{AP}=0.2$, the average cooperation level increases monotonically with rising $u$, and it remains consistently higher than that of the two-strategy scenario. In panels~(b) and (c), which $r_{AP}$ set to 0.6 and 0.8, the average cooperation level first rises and then falls as $u$ increases, and it is consistently lower than that of the two-strategy scenario. The red solid line denotes the average cooperation level in the two-strategy case involving only $ALLC$ and $ALLD$. Parameters: $Z = 60$, $N = 6$, $c = 0.1$, $b = 1$, $n_{pg} = 3$.}
\label{fig8}
\end{figure*}
\begin{figure*}
\centering
\includegraphics[width=\textwidth]{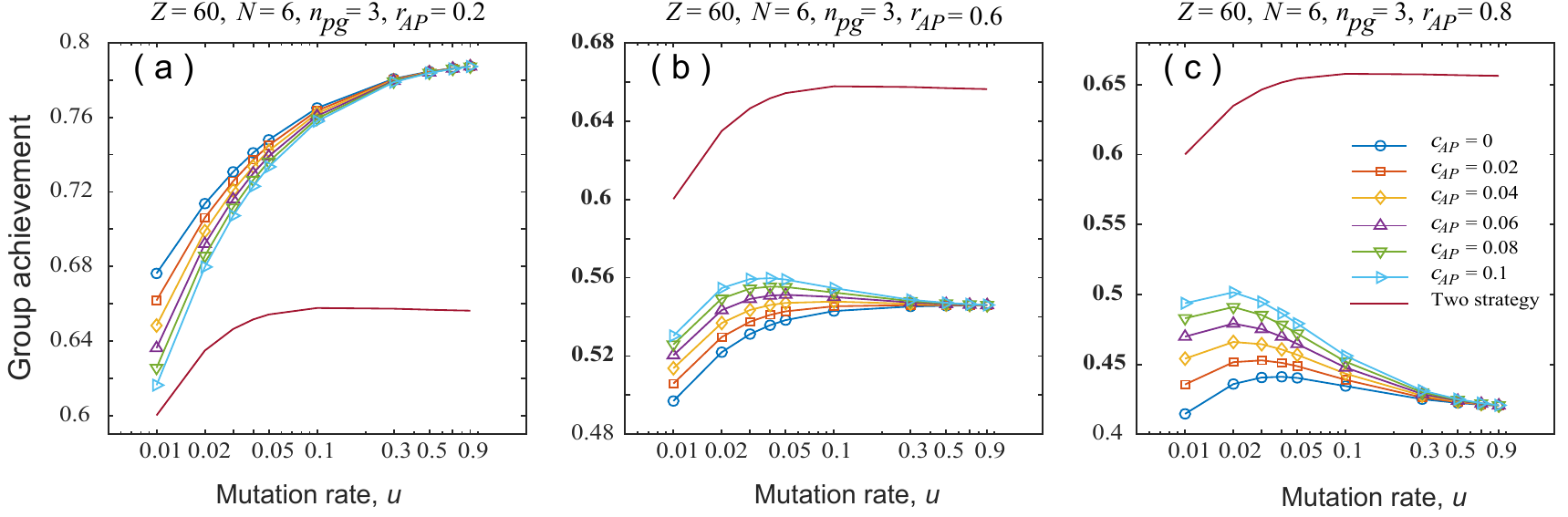}
\caption{Group achievement as a function of $u$ under distinct combinations of $c_{AP}$ and $r_{AP}$. In panel~(a), where $r_{AP}=0.2$, the group achievement of the population increases monotonically with rising $u$, and it remains consistently higher than that of the two-strategy scenario. In panels~(b) and (c), which $r_{AP}$ set to $0.6$ and $0.8$, the group achievement first rises and then falls as $u$ increases, and it is consistently lower than that of the two-strategy scenario. The red solid line denotes the group achievement in the two-strategy scenario including only $ALLC$ and $ALLD$. Parameters: $Z = 60$, $N = 6$, $c = 0.1$, $b = 1$, $n_{pg} = 3$.}
\label{fig9}
\end{figure*}

\section{Conclusion and discussion}
Researchers are increasingly using the CRD game to resolve the collective action dilemmas driven by the inherent conflicts between individual and collective interests~\cite{Milinski2008PNAS,Santos2012MMMAS,Chen2012EL,Quan2023Chaos,Terrucha2025PNAS,Wang2026PLOSCB}. Most of the previous works, however, considered that the risk of collective failure is certain during the evolutionary process. 
Nevertheless, such assumption fails to capture the inherently uncertain nature of risks observed in real-world contexts. 
A recent work has demonstrated that individuals within a population can infer variation in corresponding social environment and, more importantly, adapt their behavior accordingly~\cite{Colnaghi2023PNAS}. Motivated by this observation we here study the CRD game with uncertain risk. In this scenario, we introduce a kind of adaptive agent who pays an observation cost to infer the real-time collective failure risk, and then decides whether to defect or cooperate based on this information: adaptive agents will cooperate if the risk level exceeds their tolerance thresholds. We subsequently investigate the evolutionary dynamics of these adaptive agents and their impact on the average level of cooperation in the absence and presence of mutation.

For the no-mutation case, we find that, as the observation cost increases, the fixation probability of adaptive agents will decline, indicating that adaptive agents are conferred an evolutionary advantage only when the observation cost is relatively low. However, the fixation probability of the other two strategists will rise as the observation cost increases. Moreover, we find that as the tolerance threshold increases, the fixation probability of adaptive agents will first increase and then decrease, whereas those of the other two strategists exhibit the opposite trend. These findings suggest that the evolutionary outcomes are associated with the observation cost and tolerance threshold of adaptive agents. Furthermore, we find that the average cooperation level decreases with increasing observation costs at low tolerance thresholds, but exhibits the opposite trend at high thresholds. Our results indicate that when the tolerance threshold is low, adaptive agents are more inclined to cooperate, and in this condition, increasing the observation cost hinders the evolution of cooperative behavior. In contrast, a high tolerance threshold drives adaptive agents toward defection. In such a case, raising the observation cost curbs defective behaviors within the population, thereby fostering the evolution of cooperation. Therefore, the tolerance threshold directly shapes the behavioral preferences of adaptive agents, and a higher threshold inhibit the evolution of cooperation in the population. The average cooperation level declines as the tolerance threshold rises. Furthermore, compared to the two-strategy scenario, the introduction of adaptive agents best enhance the average cooperation level at low observation costs and tolerance thresholds, whereas this advantage diminishes as both parameters increase.  

In the presence of mutation, our results reveal that under a low tolerance threshold, both average cooperation level and group achievement decline as the observation cost rises. By contrast, when the tolerance threshold is high, these two quantities increase with rising the observation cost. Moreover, we observe that an increase in the tolerance threshold leads to a decrease in both average cooperation and group achievement. Importantly, the trends in average cooperation with respect to observation costs and tolerance thresholds remain consistent regardless of mutation, confirming the robustness of our results to variation in this parameter. Our findings suggest that under low tolerance thresholds, an increase in mutation rate markedly improves both the average cooperation level and group achievement. In this regime, lower observation costs yield higher average cooperation level and group achievement. As the tolerance threshold rises, however, these quantities exhibit a non-monotonic trend, peaking and subsequently declining with increasing mutation. Additionally, under high mutation rates, these two quantities are influenced by tolerance threshold, where a higher tolerance threshold leads to a lower converged value. Moreover, under high tolerance thresholds, higher observation costs will enhance the average cooperation level and group achievement in the population. This finding is consistent with the results drawn from cases without mutation or with low mutation probabilities. In contrast to 
the two-strategy cases, introducing adaptive agents can enhance the average cooperation level and group achievement in the population at low tolerance thresholds. Nevertheless, as the tolerance threshold increases, the introduction of adaptive agents leads to a decline in both the average cooperation level and group achievement compared to the two-strategy scenarios. Furthermore, we find that as the population size increases, the average cooperation level changes only slightly under low tolerance thresholds, whereas it exhibits a modest increase under higher tolerance thresholds. In contrast, variations in group size exert a more pronounced effect on the average cooperation level. Specifically, we find that a larger group size leads to a lower average cooperation level, which is consistent with findings from previous study~\cite{Santos2011PNAS}.

In this work, we only focused on the case where the perception of risk follows a uniform distribution. Therefore, exploring how alternative distributions influence the evolutionary dynamics in a population including these three strategists constitutes a promising direction for future research. Moreover, since well-mixed populations may have limitations in capturing the complexities of real-world network, it would also
be interesting to introduce these adaptive agents into the CRD game for investigating the evolution of cooperation in structured populations~\cite{Chen2014PRE,Allen2017Nature}. In addition, we assume that adaptive agents incur an observation cost to acquire the risk level. Nevertheless, because practical risk evaluations are inherently susceptible to measurement error, future research should systematically examine how such inaccuracies influence the evolutionary dynamics of cooperation.

\begin{acknowledgments}
This research was supported by the National Natural Science Foundation of China (Grant No. 62473081) and by the National Research, Development and Innovation Office (NKFIH) under Grant no. K142948.
\end{acknowledgments}

\section*{References}
\bibliography{aipsamp}

\end{document}